# Modeling nanoconfinement effects using active learning

*Javier E. Santos[1], Mohammed Mehana[2], Hao Wu[3], Masa Prodanovic[1], Michael J. Pyrcz[1],

Qinjun Kang[2], Nicholas Lubbers[4], Hari Viswanathan[2]

1. The Hildebrand Department of Petroleum and Geosystems Engineering. The University of Texas at Austin. Austin, Texas, USA.
2. Earth and Environmental Sciences Division, Los Alamos National Laboratory. Los Alamos, New Mexico, USA.
3. Department of Geosciences, Virginia Polytechnic Institute & State University, Blacksburg, Virginia, USA
4. Computer, Computational, and Statistical Sciences Division, Los Alamos National Laboratory, Los Alamos, New Mexico, USA.

## Abstract

Predicting the spatial configuration of gas molecules in nanopores of shale formations is crucial for fluid flow forecasting and hydrocarbon reserves estimation. The key challenge in these tight formations is that the majority of the pore sizes are less than 50 nm. At this scale, the fluid properties are affected by nanoconfinement effects due to the increased fluid-solid interactions. For instance, gas adsorption to the pore walls could account for up to 85% of the total hydrocarbon volume in a tight reservoir. Although there are analytical solutions that describe this phenomenon for simple geometries, they are not suitable for describing realistic pores, where surface roughness and geometric anisotropy play important roles. To describe these, molecular dynamics (MD) simulations are used since they consider fluid-solid and fluid-fluid interactions at the molecular level. However, MD simulations are computationally expensive, and are not able to simulate scales larger than a few connected nanopores. Alternatively, mesoscale simulation methods that handle larger domains with complex pore geometries (i.e. the lattice-Boltzmann method) cannot directly account for nanoconfinement effects, resulting in grossly inaccurate predictions. To incorporate these effects into larger domains with complex geometries, it is necessary to accelerate the computation. We present a method for building and training physics-based deep learning surrogate models to carry out fast and accurate predictions of molecular configurations of gas inside nanopores. Since training deep learning models requires extensive databases that are computationally expensive to create, we employ active learning (AL). AL reduces the overhead of creating comprehensive sets of high-fidelity data by determining where the model uncertainty is greatest, and running simulations on the fly to minimize it. The proposed workflow enables nanoconfinement effects to be rigorously considered at the mesoscale where complex connected sets of nanopores control key applications such as hydrocarbon recovery and $CO_2$ sequestration.

**Keywords**: convolutional neural networks, active learning, surrogate modeling, nanopores, adsorption, nanoconfinement, molecular dynamics, shale

# 1. Introduction

The demand of natural gas for energy generation, industrial processes, and domestic consumption has encouraged the production of unconventional resources, predominantly from shale formations. Methane, the principal hydrocarbon constituent in shale gas plays, is an attractive asset due to its energy density, ease of transport, and environmental advantages over resources like coal. Recent technological advances (hydraulic fracturing, slick water, horizontal drilling), have made possible the economic utilization of these resources[1,2]. Due to these, the world has witnessed a significant increase of natural gas production from shale reservoirs in the last decade. Nevertheless, accurately estimating hydrocarbon reserves and forecasting well performance are tasks that remain unsolved[3–5]. One of the main reasons is that most of the pore bodies in these formations are under 50 [nm] in size, hence, nanoconfinement effects must be integrated to properly model fluid behavior[6–8]. Molecular dynamics (MD) simulations can describe nanoconfinement effects accurately [9–11]. These simulations have shown that nanoconfinement effects affect multiple fluid properties, such as viscosity, density, critical point, and adsorbed concentration. However, the computational cost of MD does not allow modelling more than a few connected nanopores. Therefore, there is a pressing need to accelerate these simulations to allow large-scale modeling of the porous media while preserving the physical effects accompanying nanoconfinement. This capability has numerous applications in energy and environment, such as hydrocarbon recovery, $CO_2$ sequestration, and energy storage and conversion. All these applications require accurate simulations of nanoconfinement effects in order to predict and optimize the system behavior.

Recently, deep learning (DL) models [12] have shown outstanding results with surrogate models that capture the desired physics of a variety of scientific fields [13–16]. These are relevant because they can provide predictions that are orders of magnitude faster in computational speed than traditional full-physics numerical simulators. Recently, porous media literature presented several successful implementations of DL models. For example, enhancing the imaging resolution of shale formations by training a deep neural network[17]. Generating 3D realizations of binary geometries that honored the statistics of a pool of 3D X-ray domains [18,13] and using the morphology of porous materials to model the solution of the Navier-Stokes equation with DL[19].

The success of supervised DL models partially stems from the availability of these to learn from large amounts of labeled data during training[20]. DL models have shown an unprecedented ability to improve as more data becomes available, and can process inputs in their raw form [12]. The main bottleneck in developing surrogate models is that the simulation of physical processes (MD simulation in this work) is computationally intensive. An efficient approach would be to selectively sample points in space with the goal of creating a dataset to improve the model's generalization capabilities and accuracy when applied to data not used train the model. The ML framework of Active Learning (AL) attempts to overcome the bottleneck of insufficient training data by building a model that requests new data where there is no sufficient information; AL focuses on minimizing the size of dataset required by the networks via designing a self-improving system [21]. AL is of particular interest for systems where the training data comprises computer simulations, so that an automated workflow can continuously collect data and re-train the model [22]. AL reduces the computational expense of creating comprehensive high-fidelity datasets.

Therefore, the domain of molecular dynamics simulations provides an ideal problem context for AL techniques. In this paper, we show the effectiveness of using active learning as a means for bringing molecular information into mesoscale models of nanoporous media.

Contrary to traditional hydrocarbon reservoirs, 80% of the pores in shale are less than 50 [nm] in size [23]. In fact, the confined fluid properties and phase transitions are active research areas [24–27]. Several experimental observations and field reports have confirmed that the fluids deviate from the bulk behavior at this scale [26]. As a result, the actual in-situ gas volumes significantly deviate from predictions made using conventional methods (up to 2.5 times [28]). Likewise, for dynamic problems,research has shown that conventional methods greatly underestimate flow through nanopores, finding a discrepancies up to two orders of magnitude compared with the Hagen-Poiseuille solution [25]. Among the physical phenomena that occurs in shale formations, the adsorption of gas into the pore walls is one of the most relevant, because, in some cases, most of the gas exists as an adsorbed layer in the vicinity of the pore wall. Additionally, adsorbed layers reduce the volume of void space available for flow and affect the slippage of the fluids[29]. Therefore, it is imperative to investigate the methane adsorption behavior in shale plays at the appropriate scale. In this work, we aim to train a surrogate model to provide fast and accurate predictions in shale nanopores.

MD simulations have provided new insights about the confinement effect on the phase properties of hydrocarbons, showing how nanoconfinement affected condensation conditions in nanopores[30]. In the context of fluid flow, it has been reported flow enhancement under confinement due to slip flow at the solid boundary[31]. Researchers simulated $CO_2$–methane systems and showed that injecting $CO_2$ might be a feasible enhanced oil recovery technique for shale [32]. All these findings at the nanoscale are possible by simulating individual molecular interactions; however, the main obstacles to unleash the capabilities of molecular simulation are the computational cost and the scheme to upscale the microscopic insights to the continuum-scale. On the other hand, Lattice-Boltzmann Models (LBM) allow the continuum simulation of larger physical domains where complex connected sets of nanopores control fluid behavior[33]. Nonetheless, LBM needs to be informed by MD (in the form of pairwise interactions [34] to account for the nanoconfinement effects). There is no functional relationship to calculate these apriori.

In our proposed method, we build a convolutional neural network model that describes absorption under nanoconfinement in complex and realistic pore geometries. This model allows at least three orders of magnitude speed-up compared to conventional MD workflow enables the application to describe more complex domains accurately. We also show an efficient method for creating training sets via AL. Although we focus our efforts in adsorption, the workflow that we propose is applicable for other nanoconfinement effects, such as effective viscosity, density, and the shifting of critical points.

## 2. Results and Discussion

In this section, we attempt to obtain the density profile of a random pore geometry under different initial bulk densities. Figure 1 illustrates our workflow, wherein we learn the density profile from a dataset of molecular dynamics calculations on varying pore geometries. In this work, we utilize Convolutional Neural Networks (CNNs), a spatial variant of neural networks to create surrogate models for MD. CNN parameters consist of convolutional filters, which are small spatial arrays of trainable parameters (3x3 in our case). These filters are convolved across the image plane (a local dot product operation) to produce a filter map that preserves the image structure of the data. Through multiple applications of convolutional layers, the model is able to build the solution image (in our case, the MD density profile), starting from the input image (the nanopore). Specifically, the input to the model is an image where the solid boundaries and the outside of the pore are labeled with zeros and the rest of the image with the initial density of the gas. The output of the model is the density profile image, where the gas takes on a heterogeneous structure due to the formation of adsorbed layers.

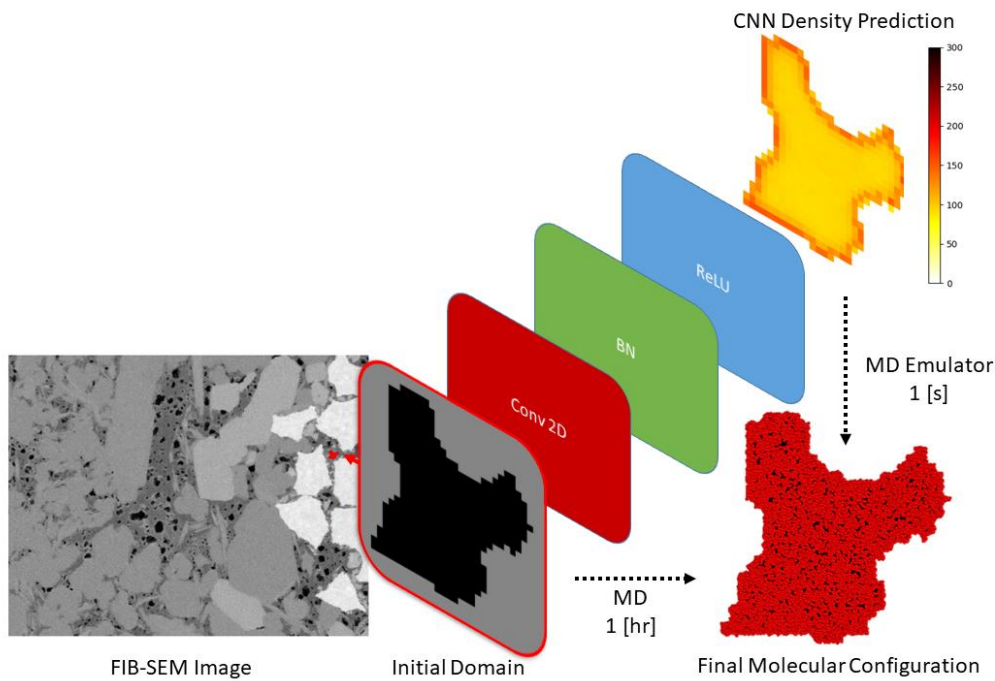

**Figure 1**: The workflow. Conventionally, starting from an initial domain (in this case, a pore from a FIB-SEM image of the Vaca Muerta formation), we generate the binarized pore geometry. In a traditional workflow, an MD simulation is carried out to obtain the final configuration of gas in the system. Each MD calculation takes on the order of one hour to complete. A CNN model learns from a database of such calculations to make predictions in less than one millisecond.

We consider pore geometries of three types: (1) Simple geometries of ellipses and rectangles, (2) Irregular synthetic geometries containing corners and 3) Discretized pores from a 3D FIB-SEM image of the Vaca Muerta formation [35]. We present individual examples of each subset in Figure 2. For more information on these datasets, see section 3.1.

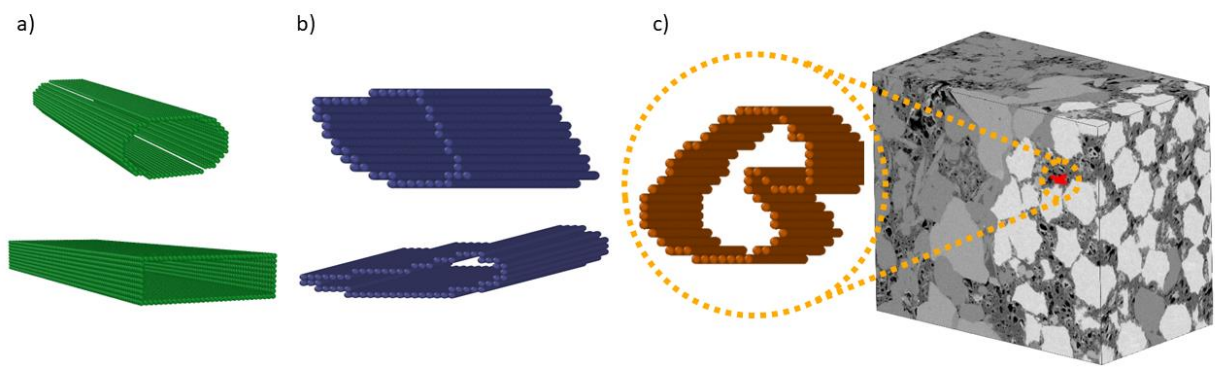

**Figure 2**: Three simulation domain subsets: a) simple geometries, b) irregular geometries, and c) real pores from a 3D image of the Vaca Muerta formation. The solid spheres show the walls of our domains (which are impermeable). The images were created using the Ovito software [36].

## 2.1 Selecting a suitable model

Since it would be impossible to have a training set with every possible geometry/gas density that could occur in nature, we use the simple shape dataset (Figure 2.a) to train different models, followed by a performance assessment of the irregular geometries (Figure 2.b). This process shows the generalization capability of each model. As a baseline, we compare deep learning models against linear regression on geometric features (see section 3.3). The fitting of the model was done using 1200 samples (70% fitting, 20% validation, and 10% test) of the simple shape dataset. We also test these models on the irregular synthetic geometries (1500 samples) to investigate how well the model can generalize to more complex data.

### 2.1.2 Comparison of CNN architectures

To assess the optimal complexity and number of trainable parameters of our convolutional neural network model, we test many network architectures; the goal of this exercise to select the best network structure. The architecture variables that we tested are the number of stacks and number of filters in the first convolutional layer (a detailed explanation of these variables is provided in Sec 3.4). We tested networks that ranged from one to nine stacks, and the number of first layer filters varying from two to sixty-four. Five identical networks are trained for each architecture using different random number seeds, and their results were averaged. The average test error of each architecture (with five trained networks) is shown in Figure 3, compared to the baseline generated by linear regression.

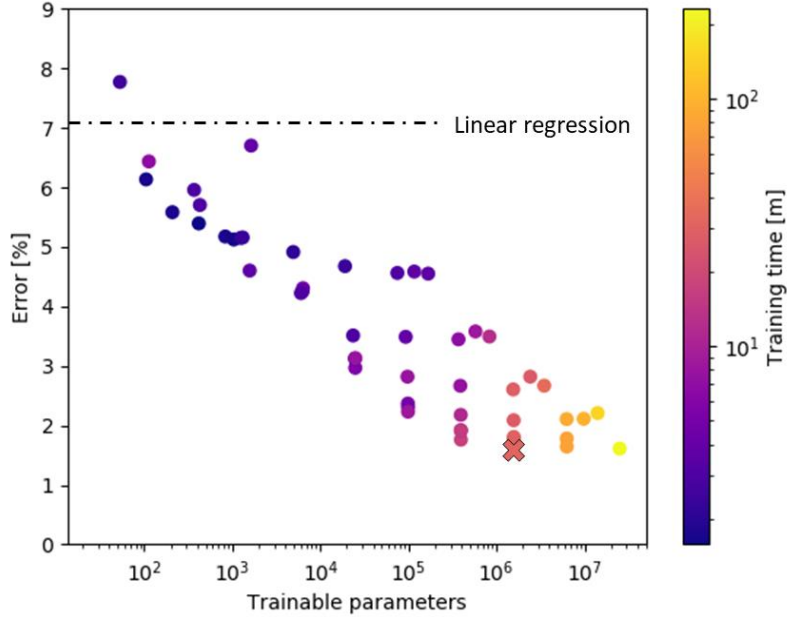

**Figure 3**. Sensitivity analysis of the network architecture parameters on the simple shapes test set, measured as mean relative error per pixel of the predictions of the five networks. The selected network is highlighted with a cross (seven stacks with eight filters in the first layer). This model yields an average error of 1.6% with a training time of 31 minutes.

We selected a network architecture (Figure 3, cross that performs accurately (1.6% relative error) in the simple shape test set while minimizing the training time of the network. The mean relative error of five trained networks of the selected architecture on the irregular shape set is 5.3% (compared to 8.8% of the best linear regression model), with a maximum individual pixel error of 96%. Note that this model was trained with just simple shapes (ellipses and rectangles); hence we do not expect high accuracy in a more complicated irregular shape dataset.

### 2.1.3 Physics-informed network improvements

Here we present physically-motivated improvements of the model architecture to improve the model performance. To ensure that there is no mass of methane leaking outside the pore walls in the prediction of the gas density of our network, we added an additional layer augmented to the end of the selected architecture (of Section 2.1.2). This layer is a mask containing the initial gas density inside the pore. The layer scales the results of the CNN by the input density, and ensures that the values of gas density at the solid walls remain to be zero. Conceptually, this can be represented as a chain of nonlinear functions $f_i$ (for each layer $i$ of the network) as follows, where $\rho_0$ represents the network input image, and $\rho_{pred}$ the predicted profile from the network:

$$\rho_{pred} = \rho_0( f_n ( f_{n-1} ( \ldots f_0 ( \rho_0 ) ) ) ) \qquad (1)$$

Using this new structure, the network learns to create a function that scales the initial homogeneous density to the final one exhibiting the adsorbed profile. The result of this addition is shown in Table 1. This improves the performance of the model; the error on irregular shapes decreases from 5.3% to 4.8%, and the total mass error is reduced from 3.5% to 2.6% with this simple additional factor.

**Table 1**. Comparison of adding a density scaling layer to the network with the mass balance error of the prediction.

|  | Test set error (from the simple shapes subset) | Irregular shapes set error | Maximum pixel error in the irregular shape set | Average total mass error in the irregular shape set |
|---|---|---|---|---|
| Standard CNN | 1.6 % | 5.3 % | 96 % | 3.5% |
| Density scaling | 1.4 % | 4.8 % | 78 % | 2.6% |

Since we average the predictions across five models, we would like to minimize the disagreement between them. To achieve this, we added a regularizing term to our loss function:

$$L = |\rho_{true} - \rho_{pred}| + \alpha |\overline{\rho_{in}} - \overline{\rho_{pred}}| \quad\quad\quad\quad (2)$$

Where $\rho_{true}$ and $\rho_{pred}$ stand for the true density (from the MD simulation), and the predicted one (from the CNN), respectively. In the second term, $\overline{\rho_{in}}$ is the mean density (the initial state, before the simulation), and α is a hyperparameter. This additional term can be seen as a penalty in case there is a difference between the initial and the final mass of gas.

A comparison of the networks with the density scaling and this proposed regularization using different α values is displayed in Table 2. We stress that these results are from models trained with simple shapes, in order to assess the best possible network architecture and hyperparameters. From Table 2, α does not have a strong effect on the performance of the ensemble of networks in aggregate, however, it does affect the disagreement between individual networks in the ensemble. We selected an α value of four since it exhibits the lowest standard deviation throughout 1500 samples.

**Table 2.** Relative error and standard deviation of the predictions (across the five networks) using different α coefficients that regularize the network training.

| α | Relative error on Irregular shape set [%] | Standard deviation of the relative errors across the five networks [%] |
|---|---|---|
| 0 | 4.8 | 0.13 |
| 1 | 4.6 | 0.15 |
| 2 | 4.6 | 0.15 |
| 3 | 4.7 | 0.13 |
| 4 | 4.6 | 0.09 |
| 5 | 4.6 | 0.14 |

## 2.2 Active Learning (AL)

Since MD calculations are relatively expensive, and the space of possible pore geometries is enormous, it is important to develop a model that generalizes to the complexity of real pore geometries from different sources without needing to know in advance what pores it will be applied to. To this end, we constructed an AL workflow to build models which generalize to the Vaca Muerta pores by training to a limited number of synthetic pore geometries,

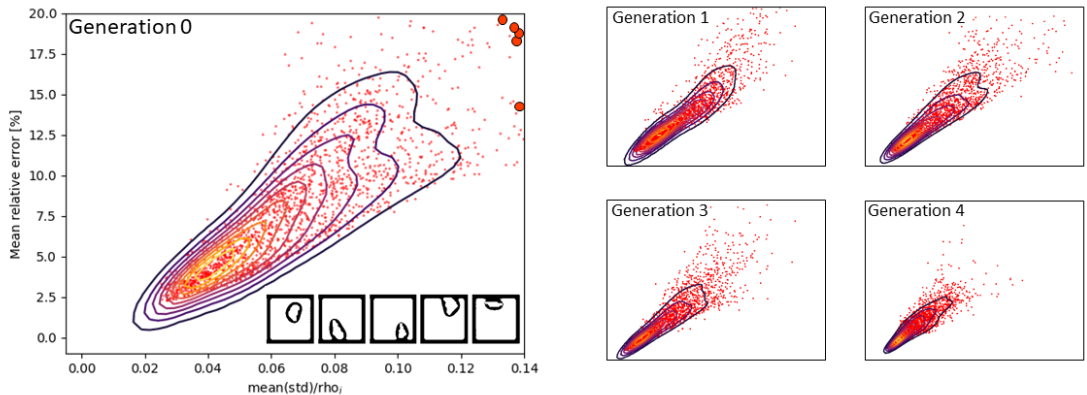

**Figure 4**: Uncertainty reduction. Mean relative error vs standard deviation of the predictions, where each data point (red dots) corresponds to a simulation case. Each generation depicts an Active Learning iteration (adding the n-least-certain examples). These n-simulations can be done on the fly, using the AL criterion. In the first plot to the left, we show the five cases (bold red dots) that yield in the highest standard deviation of the predictions. The pore shapes of these are shown in the gray squares. The five panels have the same axis limits.

The strategy adaptively augments the training set by selecting pore geometries have not been trained to but are likely to improve the model performance. The method follows the Query-by-Committee [37] technique, which takes advantage of the fact that disagreement between different models is a reasonable proxy for model uncertainty, and can be used to prioritize additions to the dataset. For further details, see section 3.5. We start from an pool of examples containing 3000 samples of simple and irregular shapes at different initial density. The initial training set is comprised of eight pores drawn from the simple shapes subset (Figure 1.a). After the training of the five networks is completed (with only these eight examples), predictions with each trained model are carried out throughout the unlabeled samples in the pool. To estimate the uncertainty of the predictions, the standard deviation of the predicted density (across the five network predictions) is recorded. The standard deviation per example is normalized using the initial density of the domain, so that the disagreement is dimensionless. Figure 4, left panel, shows strong relationship between disagreement and model error. The eight unlabeled samples exhibiting the highest value are simulated using MD, and added to the training set, then the networks are retrained using these. The process is carried out repeatedly, yielding generations of models trained to progressively larger datasets. Figure 4, right panels, shows that each progressive generation yields both lower model disagreement and lower model error.

To test the effectiveness of the Active Learning method, we randomly select 60 pores from the FIB-SEM Vaca Muerta pore dataset and run simulations varying the overall gas density, and assessed the performance of networks as they selected new data. We tested three different strategies: adding the samples from the pool with the highest mean value (across all the pixels in one image) of standard deviation, adding the samples with the highest maximum value (at a single location), and, for comparison, adding samples randomly, with no active selection strategy.

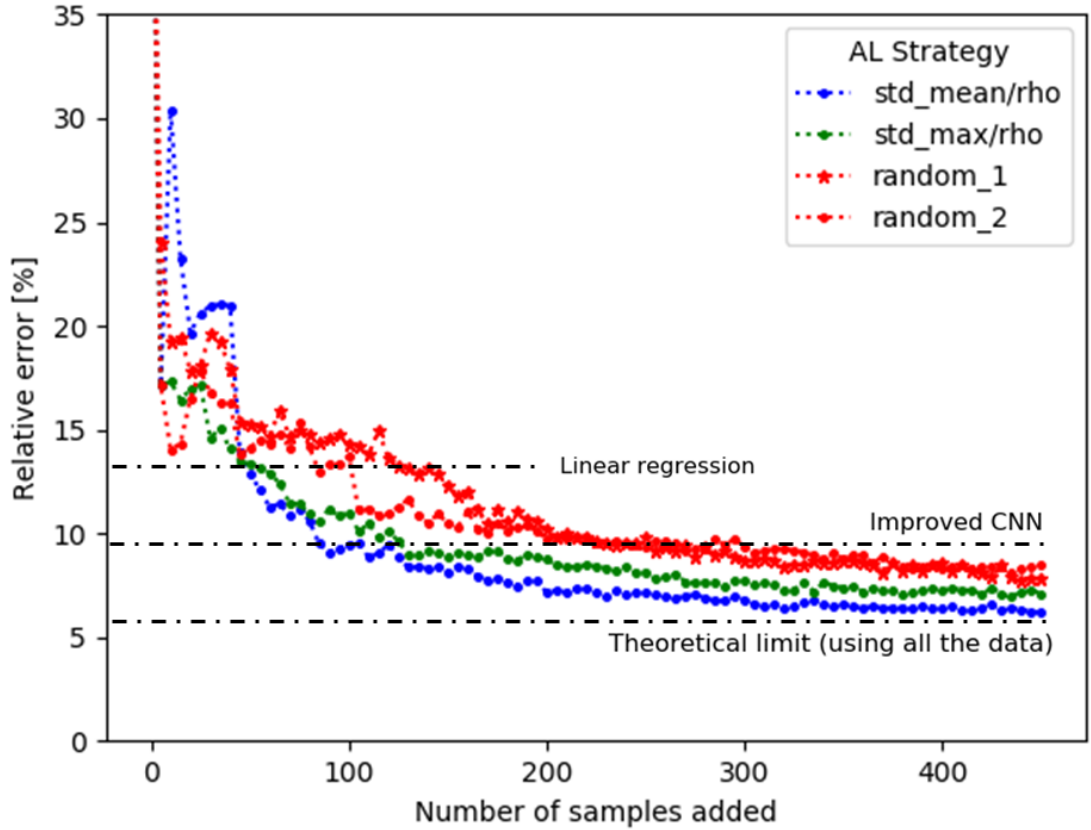

**Figure 5**: Performance of the different active learning strategies over the Vaca Muerta set. Each AL generation is defined with a marker. Two random sampling runs with different seeds were carried out to show the superiority of the other active learning strategies. In black dashed lines, the results of the linear regression and the CNN model of Section 2.1.3 both trained with 1500 examples, is shown. The theoretical limit is a network trained with all of the data available (3000 examples of simple and irregular synthetic geometries).

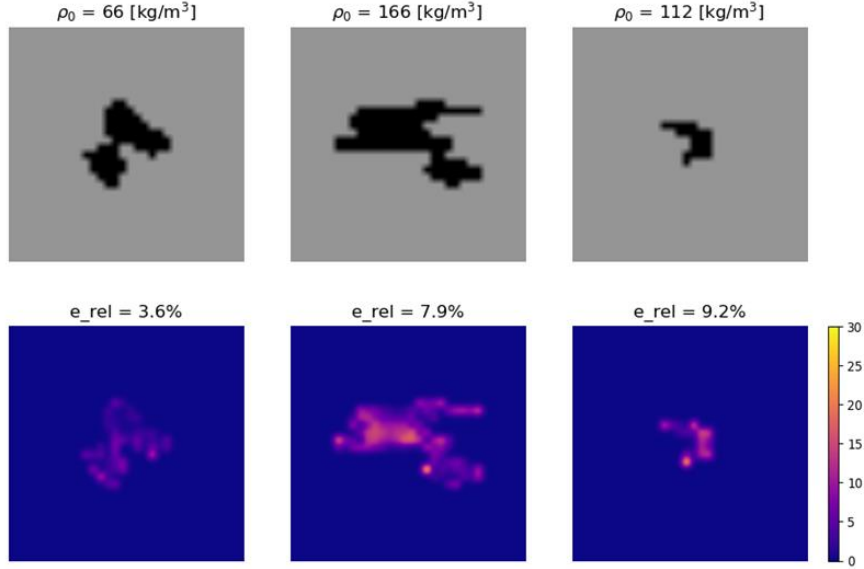

**Figure 6:** Relative error of the predictions of the Vaca Muerta test set pores using the model trained with AL. In this figure, we show a pore with low, medium and high average relative error. The top panels show the pore geometries, and the bottom panels show the error of the network prediction. The color scheme indicates pixel-wise percentage error.

The results of 90 AL generations (which are defined as every time when more examples are added and the networks retrained) are shown in Figure 5. Dashed lines show comparisons with other methods: linear regression, the CNN trained to all of the simple shapes dataset, and a CNN trained to all of the simple and irregular synthetic shapes (that is, the entire pool of samples available during active learning), From Figure 5, we can see that a targeted training set containing only 90 examples selected with the AL technique outperforms the simple-shape CNN model that is trained 1200 examples. The random sampling approach needs almost three times the amount of data to perform similarly. The curves plateau due to the fact that the pool of candidate samples is limited, but the active selection strategies plateau much more quickly than the random selection strategy. This shows that the AL strategy is more robust and predictive in generalizing to a new, more complex distribution of pore shapes. Figure 6 presents the relative error of the best active learning model for three pores that are examples of low, typical, and high error among the Vaca Meurta pore dataset.

## 2.3 Conclusions and Future Work

In this paper, we build machine learning-based surrogate models for predicting the molecular density profile of a gas-filled pore under confinement. We present a methodology to select the optimal depth and size of a CNN architecture and tune the hyperparameters to carry out physics-based predictions in images. Moreover, we demonstrate additional enhancements to the model to obtain better predictions: by applying physical considerations to the network, we ensure that the predictions we obtain have a physical basis. Finally, we demonstrate an effective way of building datasets with Active Learning to train robust predictive models efficiently. By using a query-by-committee strategy, we are able to achieve generality with limited data, only needing simulations to fill targeted areas of the data space where uncertainty is large. The method that we present is able to provide predictions of methane density in nanopores in three milliseconds (when using five networks) compared to close to an hour with MD, yielding speedups factors on the order of 1 million. We show how an active learning strategy gets the same accuracy of a network trained with a comprehensive dataset (3 000 simulations) using only 15% of this data. By doing this, it is possible to predict the density profile from an SEM image of a specific formation to enhance the current way of estimating gas reserves.

Adsorption is only one example of the nanoconfinement effects relevant in shale. In this nanoscale regime, all traditional computational fluid dynamics approaches (e.g. LBM) break as a result of being defined in the continuum limit, and more expensive molecular-level calculations must be performed for physics-based modeling. Using our workflow, the expense of these molecular calculations can be amortized and prioritized using a machine learning model that can be rapidly deployed to new pores.

While having a fast proxy for molecular dynamics in nanopores has value in itself e.g. for estimating gas reserves, we would like to point towards what we believe to be an important future application of our methods: The use of the surrogate for the parameterization of continuum models of nanoscale porous media. A fast surrogate would enable the upscaling of nanoconfinement effects into e.g. an extended LBM formulation that can treat fluid flow as a coupled problem between many pores of spanning many length scales. Such a scale-bridging model would enable a fast computational method to address the role of nanoconfinement in porous media for a variety of applications in engineering, energy sciences, and environmental sciences.

# 3. Methods

In this section, we present the techniques that we utilize in our workflow. We first describe several sources of nanopore geometries, which consist of three subsets: simple synthetic geometries, complex synthetic geometries, and empirically obtained geometries. From these subsets, we use Molecular Dynamics to obtain fluid density distributions under nanoconfinement. We model these densities with Convolutional Neural Network models, which naturally capture the spatial correlations of the data. Finally, we describe our Active Learning workflow, which is powered by an ensemble query-by-committee algorithm [37] to efficiently select new data to augment the training set and retrain the models as needed.

## 3.1 Nanopore simulation domains

To create a comprehensive dataset for training and testing the models, we produced three different domain subsets with increasing complexity. All of them are bounded in a 120x120x373 [Å] box with periodic boundary conditions in all directions. These examples are 3D geometries with a constant cross section in the Z-axis, forming a nanotube. The motivation behind having three different datasets is to be able to assess how the different models perform with increasingly intricate domains.

The three types of nanopores considered here are the following:

- Simple geometries: This subset contains ellipses and rectangles of varying cross-sectional area, centered in different locations across the plane (Figure 1.a). These types of geometries are amenable to analytical solutions.
- Irregular geometries: synthetic geometries with different number of corners, concave and convex shapes, and varying cross-sectional areas (Figure 1.b).
- Vaca Muerta pores: Discretized pores from an 3D FIB-SEM image of the Vaca Muerta formation in Argentina [35] (Figure 1.c). This image is obtained using a scanning electron microscope (SEM) combined with a focused ion beam (FIB). The setup enables to image volumes, which fully describe minerals, pore shapes, and connectivity. Using this imaging technique, it is possible to capture inter and intra-particle pores, and organic matter pores that typically are in the nanometer range.

The diversity of shapes in each subset can be quantified via the shape factor G, which is expressed as the ratio of the cross-sectional area A to the square of the perimeter p (Mason and Morrow, 1991):

$$G = \frac{A}{p^2}$$ ……………………………………………………………………….. (3)

The shape factor of the three different data subsets is illustrated in Figure 7. There we can see that the simple geometries have a bimodal distribution due to the two geometric shapes that constitute this set. The real pores have the widest shape factor distribution, meaning that these pores have cross-sections that change dramatically in shape.

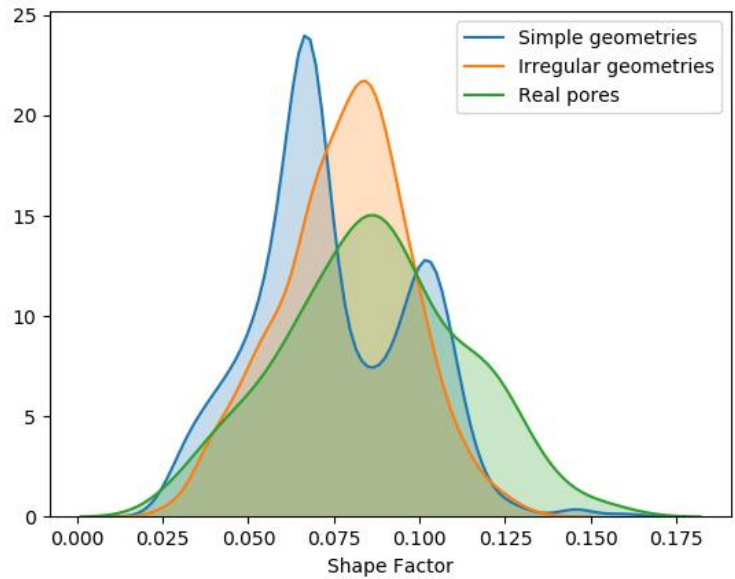

**Figure 7**: Shape factor (Equation 1) distributions of the three different data subsets.

Our goal is to train a model to perform accurately in real pores using only synthetic geometries for training (without leaking any information of this subset's characteristics apriori) while minimizing the amount of training data needed for this task.

3.2 Molecular dynamics (MD) simulation

Since our target pores are nanometer-size, continuum and mesoscopic approaches fail to capture the relevant physics at this scale. Therefore, we utilize molecular dynamics to simulate the system of interest. We chose the LAMMPS software [38] to carry out our simulations. Specifically, we aim to estimate the density profile of a fluid inside a pore under confinement conditions. The density profile will be significantly affected by liquid-solid interactions at the pore wall, especially for pores less than 50 nm. It should be noted that, methane adsorption is not a critical process that affects hydrocarbon extraction for larger pores (i.e. sandstones). In fact, methane adsorption can often be neglected in these cases. However, for smaller pores, due to the large surface to volume ratio, gas adsorbed to the solid pore wall could account for up to 85% of the total hydrocarbon volume in a pore, acting as a major factor affecting production.

We used frozen methane molecules to simulate the solid walls of our structures. The gas inside the pore is assumed to follow the united-atom model at a given initial density (controlled by the number of molecules placed inside the domain). We used the TraPPE force field [39] to describe the molecular interactions with a cut-off range of 10 [Å]. To simulate accurately the adsorption phenomenon at the nanoscale, the equilibrium of the system must be ensured. To accomplish this, we start the simulation with an energy-minimization stage that avoids the overlapping of atoms, followed by an equilibration stage where a canonical ensemble is used for 500 [ps] to bring the system to the equilibrium state. Finally, we carry out a sampling phase for one [ns] in a constant temperature ensemble using a Nose-Hoover thermostat. Figure 8 shows the evolution of the molecular profile of a nanopore extracted from the SEM image of the Vaca Muerta formation through these stages. During the equilibration, the system structure is arranged where the fluid molecules adhere to the solid interface, which macroscopically translates to the adsorption phenomenon.

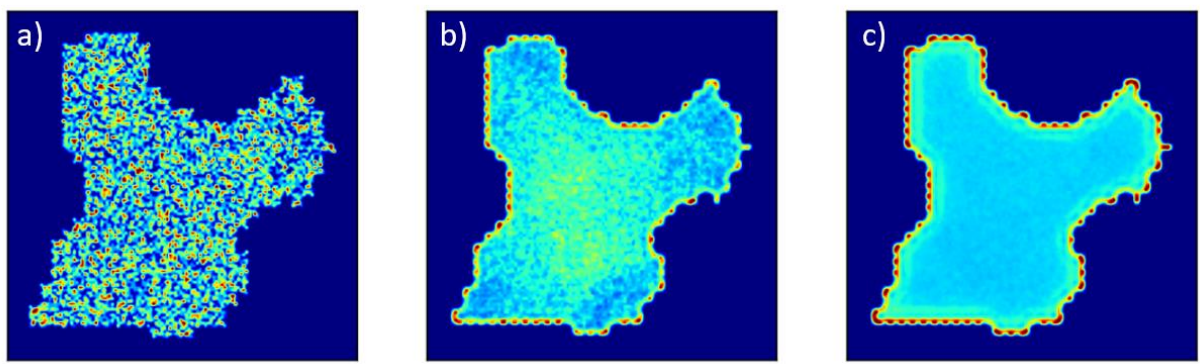

**Figure 8**: Steps of our simulation workflow showing the evolution in position of the gas molecules. a) shows the initial domain (homogeneous gas density), b) shows the density profile after 500 [ps], and c) shows the density profile after equilibrium. These are in arbitrary units; we aim to show how the molecules are preferentially located near the solid boundary.

To create a diverse data set of different pore geometries and methane density, a random number of molecules were distributed inside the nanopores (over a range of average density between 50 and 250 [kg/m3]). After the simulation run (with computational time of about one hour in a desktop computer), we average the gas density in the Z-coordinate to calculate a representative density profile. Finally, we bin these results into a 32x32 mesh (with a resolution of 3.75 [Å] per bin, slightly larger than the effective diameter of the methane molecule).

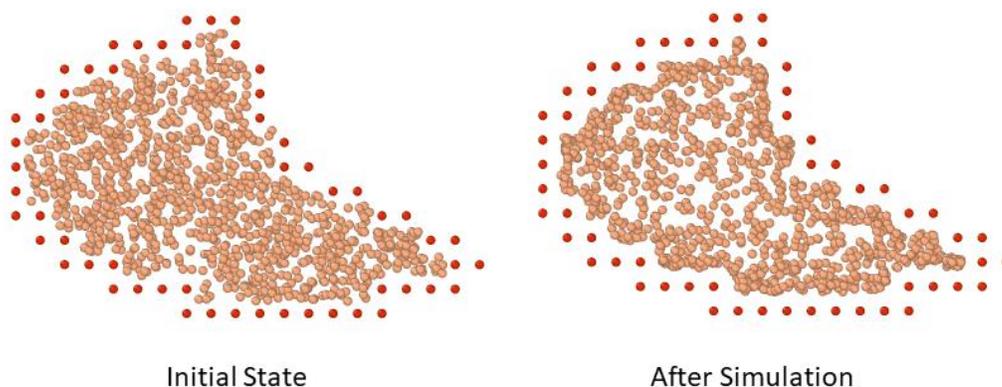

**Figure 9**: Cross-sectional view of one sample of the irregular geometry subset. The left figure shows the initial state of the system (before the canonical ensemble simulation), where the free gas molecules (light brown), are placed homogeneously in the space bounded by the impermeable pore walls (red). The average density of this system is 50 [kg/m3]. The figure to the right shows the converged state after the MD simulation. The amount of gas molecules (density) near the walls increases due to the adsorption phenomenon (220% higher than the bulk density for this case).

In the simulation results, we observe a heterogeneous density profile across the pore, where an adsorbed layer circumscribes the solid walls (Figure 9). We emphasize that for irregularly shaped nanopores, the adsorption profile is complex, and greatly influences the hydrocarbon transport. The molecules in this layer are held by the Van der Waals interactions between the solid walls and the fluid molecules. We observed a gradual decrease in the methane density as we move from the adsorbed layer to the bulk fluid (pore center). This behavior is strongly influenced by the pore geometry, and the initial density of the gas. To highlight this, we carried out four simulations at different initial density in the same domain. Figure 10 shows the big impact that geometry and initial density have on the adsorption profile. As the initial density increases, the adsorption peaks diminish. Regarding the pore geometry, where fluid molecules interact with multiple solid surfaces, they form higher density clusters.

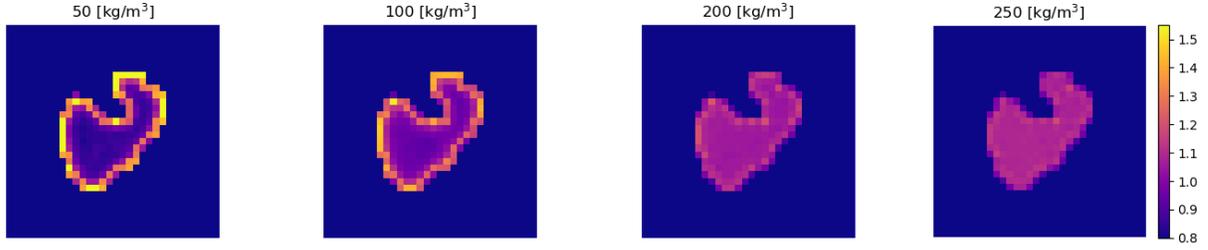

**Figure 10**: Effect of geometry and initial density in the final solution. The plots show the results of four simulations carried out in the same domain using a different initial density (specified on top of each image). We normalized the results using the initial density of each domain.

### 3.3 Linear Regression

Linear regression is a simple model that trains very fast, and provides a highly interpretable baseline to assess machine learning models with more parameters. Linear regression models a scalar relationship between distinct input features and an output response (in this case, gas density) by assigning a weight to each of the input variables, which correspond to how the function changes with the value of the input data. These models are easy to fit, and can provide a reference point to compare the performance of our deep learning models. The features that we use to train the linear models are the initial density ($\rho_0$), Euclidean distance ($E_{dist}$) and the sample porosity ($\Phi$). The Euclidean distance measures the separation between a point inside a pore with the closest wall. The porosity is obtained by dividing the area of the pore over the total cross-sectional area.

**Table 3**: Results of the linear regression models using different input features.

| | Inputs | Test set error (20% of the simple shapes) | Irregular shapes set error (1500 samples) | Resultant equation |
|---|---|---|---|---|
| 1 | Initial density | 9.01% | 10.65% | $\rho = 1.0035*\rho_0$ |
| 2 | Initial density and Euclidean distance | 7.41% | 10.50% | $\rho = 0.998*\rho_0 - 4.17*E_{dist}$ |
| 3 | Initial density, Euclidean distance, and porosity | 7.04% | 8.83% | $\rho = 1.0034*\rho_0 - 4.93*E_{dist} + 0.2772*\Phi$ |

Since the systems do not present strong density contrasts (it is always within the same order of magnitude), the initial density gets a weight very close to one in every case (see results in Table 1). This is due to the fact that mass is conserved in the MD simulation, so a weight close to one minimizes the least squares regression. This leads to good predictions in the center of big pores, where the accuracy decreases near the walls (as seen in Figure 11).

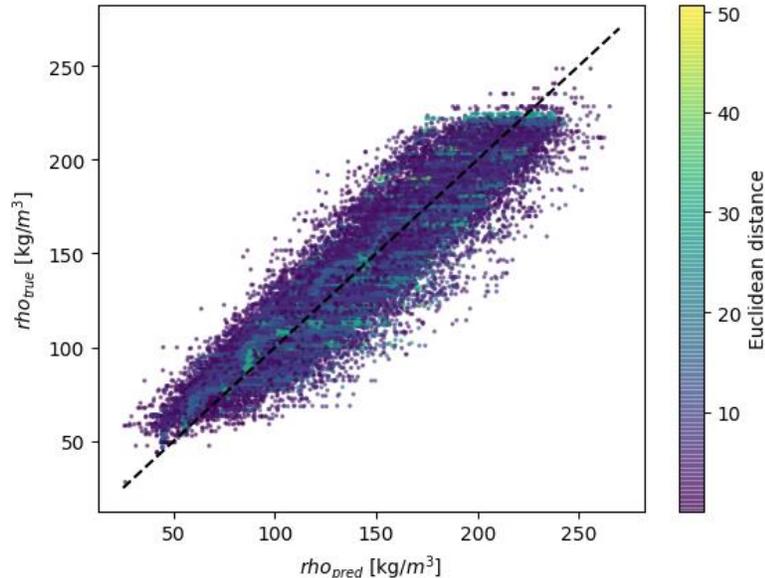

**Figure 11**: True vs the predicted density per pixel on the test set using the third linear regression model (Table 1 line 3). The color represents the distance from the solid walls, from this we can see that the model performs better when it is far away from the walls where the adsorption effects are almost negligible.

### 3.4 Convolutional Neural Networks

Since the pore geometry and initial density determine the final distribution of fluid molecules, we propose employing convolutional neural networks (CNN) to create surrogate models that capture the impact of these spatial features. In this subsection, we will first provide an overview of CNNs. Then, we describe the details of our network architecture, hyperparameters, and initialization scheme to create predictive machine learning models for estimating the density distribution inside a pore. This workflow is agnostic to the application, and could be adopted to predict other properties in porous media (either in the solid or through the void). In this paper, we work with 2D images that can fit an entire closed system (pore with impermeable walls) to present our methodology, since adsorption happens locally. 3D open systems can be handled with the addition of inputs that inform the model about the local and global boundary conditions of the system [13].

CNNs have excelled in the field of computer vision outperforming classical machine learning methods [20,40]. These models have shown a remarkable capacity to identify complex spatial relations in many datasets. By utilizing the discrete convolution operation instead of a dense matrix multiplication (i.e. a fully connected feed-forward network), convolutional layers process local spatial relationships (sparse interactions) across the domain. CNNs utilize filters that are much smaller than the input image (in our case 3 by 3 pixels), that extract general, and meaningful information about the domain in an efficient manner. By stacking convolutional layers, the network extracts features at different levels of abstraction with an increasingly wider receptive field. Finally, the convolution layers are equivariant to translation, which means that if the input feature is shifted, their output will be shifted by the same distance (by creating, in this case, a 2D feature map). This means that the exact location of the pore within the plane does not matter. This is particularly useful in pattern recognition, because CNNs allow inputs of variable size. Using this structure, a network can be trained to learn complex, non-linear relationships between inputs and outputs using the backpropagation algorithm.

Our aim is to create a function that is able to map a pore geometry initially filled with gas at a certain, homogeneous density, with the density profile at equilibrium (as provided by molecular dynamics simulation). To build this relationship, we tested different convolutional neural network architectures to create a model that is able to take in account the pore geometry and density to make accurate predictions.

Since the density fields that are obtained using the MD simulations are composed mainly by high-level features, we selected an architecture that is structured by stacked layers that do not pool (reduce its original size) the image, preserving its dimensions at every stage of the process. The network architecture consists of n-stacks of: a 2D convolutional, a batch normalization, and a ReLu activation layer (as depicted in Figure 1). The first convolutional layer is assigned a set number of filters. The subsequent convolutional layers have twice as many filters as their predecessors (as demonstrated in [41]) The last layer of the network is a Softplus, which adds flexibility to carryout density predictions outside the training range (extrapolating). The networks are implemented in Pytorch 1.0 [42]. For computational efficiency, the network is trained using single precision.

Parting from the desired CNN architecture (this process is explained in detail in Section 3.1.2), the training process is carried out as follows:

First, we initialize the network weights using the Kaiming criterion[43]. This method increases the mean and the standard deviation of each layer's initial weights to prevent the output of the activation functions from either exploding or vanishing, which would yield in the gradient not flowing during the training process (resulting in the network not learning the relationship between pore shape and gas density). This is shown to help in the gradient propagation for very deep networks by taking into consideration the specific activation functions of the model.

An initial learning rate of 1e-3 is set, where the weights are optimized using Adam[44] with a batch size of 32 random samples. We selected the absolute value (L1) as our loss function due to the strong presence of outliers (adsorbed layer). We use a plateau-based learning rate scheduler with a patience of 50: At any time, if 50 epochs have passed, and the validation loss has not lowered by at least 1, the learning rate is multiplied by a factor of 0.1. If the validation loss does not decrease by 1% for 100 consecutive epochs, the training process is stopped. We use early-stopping; the final network parameters taken are those that had the minimum validation loss over the entire training period. Our network converges in approximately 300 epochs (the training times are shown in Figure 3).

## 3.5 Active Learning

Next, we describe our active learning strategy to minimize the number of nano-scale simulations required as part of the model training. CNNs have demonstrated great performance in pattern recognition tasks when very large labeled datasets are provided. Nevertheless, creating comprehensive training sets, especially for physics-based simulations, is an expensive task. For example, to create a set of 1000 labeled pairs of MD data, one would need roughly one month of computational time on a desktop computer. Due to this hurdle, we employ an active learning method to label our data in an efficient manner. AL selects images (candidates for training) from a big pool of data to label them (in our case, they are labelled by performing an MD simulation) with the goal of minimizing the relative error of the model.

Parting from the selected network architecture, we trained five identical neural networks that differ only in the initial random number used (which only affects the network initialization, and the order of the samples employed per batch of data). The first generation of five networks use the exact same dataset to train. At the end of the training process (as explained in Section 2.3), predictions with each trained model are carried out throughout the unlabeled pool of images (where only the cross-sectional shape and the initial density are provided). Since the target result (MD simulation) is unknown for this set, we record the disagreement between network predictions. If this disagreement is high, we flag the sample as candidate for labeling (this process is depicted in Figure 12). We then select the *n* samples presenting the highest disagreement, label them (by performing an MD simulation), and use them to train the next generation of networks. By adding these examples to the initial training set, we are reducing the uncertainty space of the models (Figure 4).

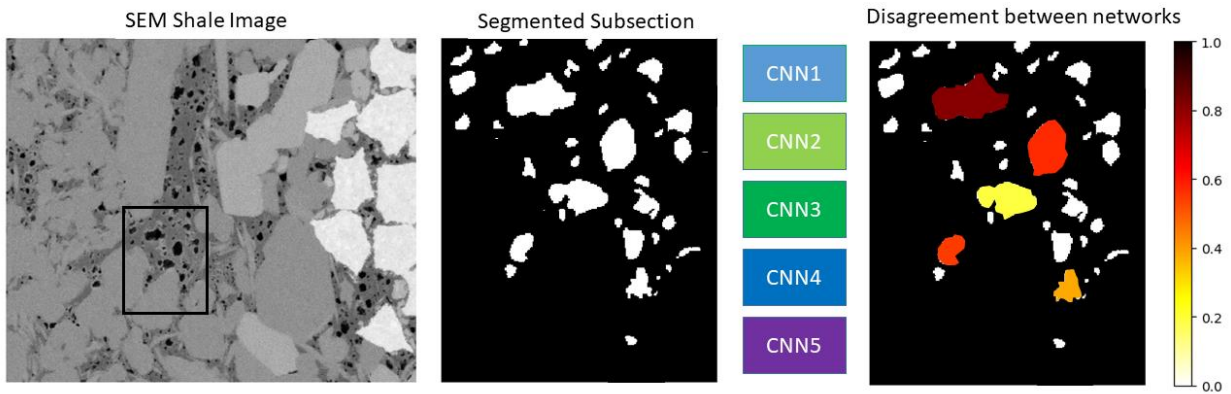

**Figure 12**: Active learning workflow. We use five trained neural networks (shown as boxes) to make predictions in pores with different shapes. We then flag the locations where these five networks diverge the most. The color bar is in arbitrary units.

In the results section we show comparisons with simpler models, we show how we carried the optimal architecture selection and hyper-parameter tuning, and the effectiveness of the proposed active learning method.

# Reproducibility

All the code and dataser will be published to the author's github (je-santos) upon publication.

# Acknowledgements

This work was funded in part by the U.S. Department of Energy through Los Alamos National Laboratory's Laboratory Directed Research and Development program (LANL-LDRD) and the Center for Nonlinear Studies (CNLS) at LANL. Computations were performed in part on the Darwin cluster at LANL, and the authors thank the Darwin team for their support.